\documentclass[aps,prb,twocolumn,groupedaddress,showpacs,superscriptaddress]{revtex4-1}

\usepackage[utf8]{inputenc}
\usepackage{graphicx}
\usepackage{amsmath}
\usepackage{amssymb}
\usepackage{color}
\usepackage{bm}

\DeclareMathOperator{\sech}{sech}
\DeclareMathOperator{\csch}{csch}

\def \Do {\partial}

\def \T {\mathcal{T}}

\begin{document}

\title{Screening cloud and non-Fermi-liquid scattering in topological Kondo devices}

\author{A. Latief}
\affiliation{School of Physics \& Astronomy, University of Birmingham, B15 2TT, United Kingdom}

\author{B. B\'eri}
\affiliation{School of Physics \& Astronomy, University of Birmingham, B15 2TT, United Kingdom}
\affiliation{Cavendish Laboratory \& DAMTP, University of Cambridge, CB3 0HE, United Kingdom}

\date{March 2018}

\begin{abstract}
The topological Kondo effect arises when conduction electrons in metallic leads are coupled to a mesoscopic superconducting island with Majorana fermions. Working with its minimal setup, we study the lead electron local tunneling density of states in its thermally smeared form motivated by scanning tunneling microscopy, focusing on the component $\rho_{2 k_F}$ oscillating at twice the Fermi wavenumber. As a function of temperature $T$ and at zero bias, we find that the amplitude of $\rho_{2 k_F}$ is nonmonotonic,  whereby with decreasing $T$ an exponential thermal-length-controlled increase, potentially through an intermediate Kondo logarithm, crosses over to a $T^{1/3}$ decay. The Kondo logarithm is present only for tip-junction distances sufficiently smaller than the Kondo length, thus providing information on the Kondo screening cloud. 
The low temperature decay indicates non-Fermi-liquid scattering, in particular the complete suppression of single-particle scattering at the topological Kondo fixed point. For temperatures much below the Kondo temperature, we find that the $\rho_{2 k_F}$ amplitude can be described as a universal scaling function indicative of strong correlations. In a more general context, our considerations point towards the utility of $\rho_{2 k_F}$ in studying quantum impurity systems, including extracting information about the single-particle scattering matrix. 
\end{abstract}

\maketitle

\section{Introduction}

Realizing Majorana fermions in condensed matter is a subject of intensive ongoing efforts,\cite{Wilczek-majorana-2009,Alicea-new-2012,Beenakker-search-2013} in part motivated by the potential Majorana fermions present for implementing schemes for quantum computation.\cite{Kitaev-fault-2003,Nayak-et-al-non-abelian-2008,Stern-non-abelian-2010,Alicea-majorana-2010,Oreg-et-al-helical-2010} Most ongoing studies focus on effectively one-dimensional settings where Majorana modes appear as zero-energy end states.\cite{Kitaev-unpaired-2001,Oreg-et-al-helical-2010,Lutchyn-et-al-majorana-2010,Leijnse-Flensberg-introduction-2012} Experimental candidate systems include semiconducting nanowires with strong spin-orbit coupling that are in contact with $s$-wave superconductors,\cite{Mourik-et-al-signatures-2012,Albrecht-et-al-exponential-2016,Zhang-Quantized-Majcond} nanowires formed by  ferromagnetic atomic chains that are in contact with superconductors with strong spin-orbit coupling,\cite{Nadj-Perge-et-al-proposal-2013,Nadj-Perge-et-al-observation-2014} and more recently systems based on two-dimensional electron gases.\cite{kjaergaard2016quantized,Nichele-scaling-2017}
The majority of experiments so far\cite{Mourik-et-al-signatures-2012,Das-et-al-zero-2012,Deng-et-al-anomalous-2012,Finck-et-al-anomalous-2013,Nichele-scaling-2017,Zhang-Quantized-Majcond}
focus on demonstrating the zero-energy end-state nature of Majorana modes through observing the zero bias conductance peak related features\cite{Law-et-al-majorana-2009,Flensberg-tunneling-2010,Sau-et-al-non-abelian-2010,Wimmer-et-al-quantum-2011,Das-et-al-zero-2012,Deng-et-al-anomalous-2012,ZazunovPRB16}
for Majorana-assisted tunneling into a superconducting reservoir.

Partly due to the potential alternative explanations behind the zero bias peak\cite{Bagrets-Altland-class-2012,Pikulin-et-al-zero-2012,Liu-et-al-zero-2012,Kells-et-al-near-2012,Rainis-et-al-towards-2013} and partly due to its fundamental importance and relevance to quantum computation, a key challenge is the demonstration of Majorana nonlocality.\cite{Nayak-et-al-non-abelian-2008,Hasan-Kane-topological-2010,Qi-Zhang-topological-2011}
A promising direction uses a so-called Majorana island, a Majorana device based on a superconducting island with large charging energy,\cite{Fu-electron-2010} as in a recent experiment\cite{Albrecht-et-al-exponential-2016} showing the first (though not yet definitive) signatures of electron teleportation.\cite{Fu-electron-2010}

\begin{figure}[b]
    \centering
    \includegraphics[width=0.5\textwidth]{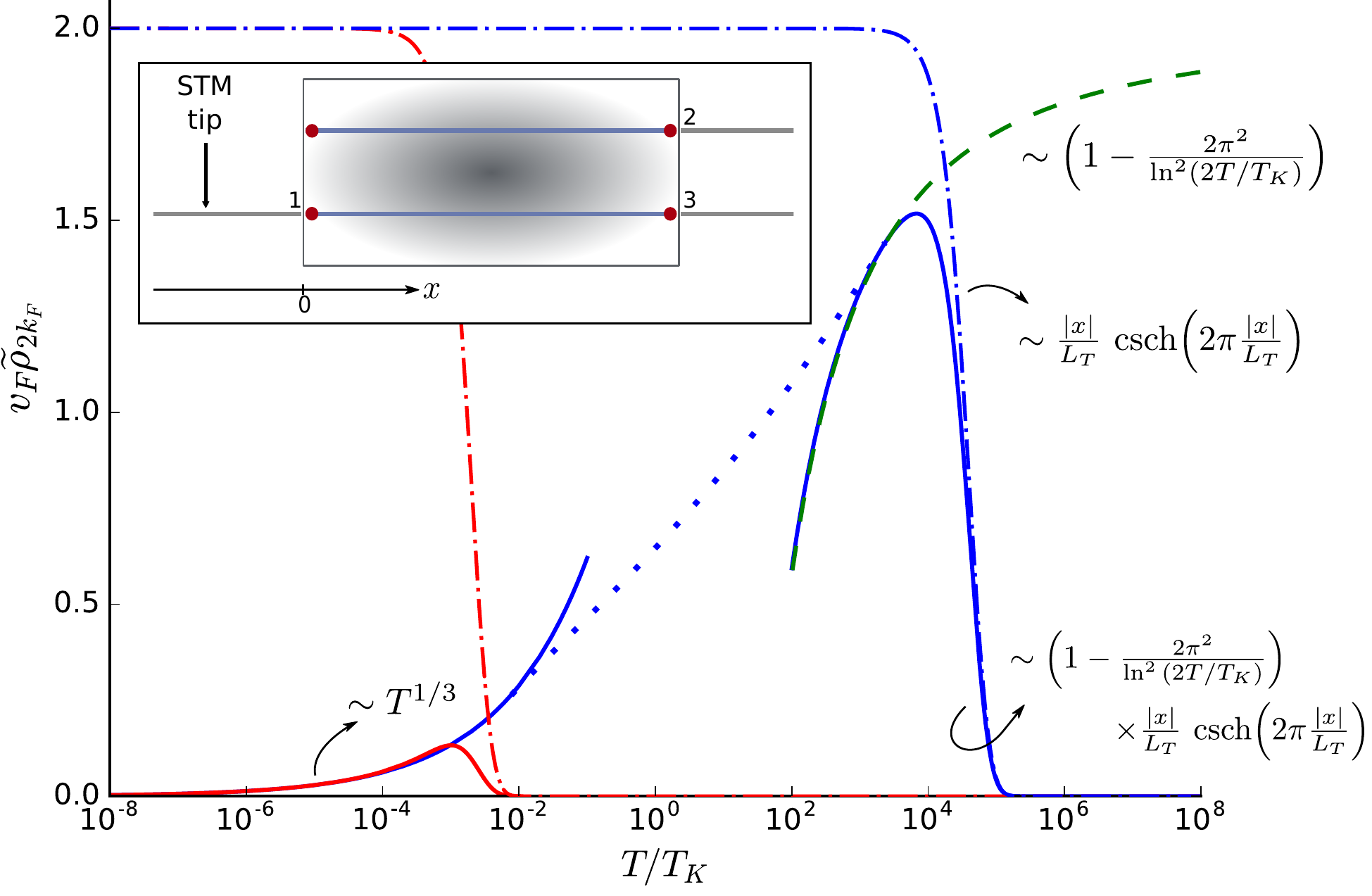}
    \caption{The amplitude $\widetilde{\rho}_{2 k_F}$ of the oscillating component of the tDOS for zero bias, for the minimal topological Kondo setup (inset) of three Majorana modes (red dots) coupled to conduction-electron leads. The tDOS may be measured using STM. In the expressions shown and below, $v_F$ is Fermi velocity, $T_K$ is Kondo temperature, and $L_K = {v_F}/{T_K}$ ($L_T = {v_F}/T$) is the Kondo length (thermal length). The solid, dashed, and dashed-dotted lines represent calculations; the dotted line extrapolates between the regions where our asymptotic approach is expected to hold. 
    For $\widetilde{\rho}_{2 k_F}(|x| \ll L_K)$ (solid blue and dots), a Kondo logarithmic peak is visible between the $T \gg T_K$ free-fermion (dash-dotted) tail and the $T \ll T_K$ power law. For $\widetilde{\rho}_{2 k_F}(|x| \gg L_K)$ (solid red), the logarithmic regime is suppressed. The dashed line is $\widetilde{\rho}_{2 k_F}(x=0)$. }
    \label{fig-setup}
\end{figure}

A compelling signature would be provided by the so-called topological Kondo effect,\cite{Beri-Cooper-topological-2012,Altland-Egger-multiterminal-2013,Beri-majorana-2013,Galpin-et-al-conductance-2014,Altland-et-al-bethe-2014,Altland-et-al-multichannel-2014,Zazunov-et-al-transport-2014,ErikssonPRL14,Eriksson-et-al-tunneling-2014,Plugge-et-al-kondo-2016,HerviouPRB16,ZazunovPRL17,Beri-exact-2017,MichaeliPRB17,Gau2018} for which devices could be constructed with only a moderate increase in complexity compared to those under current investigation. Specifically, the minimal configuration consists of a Majorana island connected to three leads of noninteracting conduction electrons via three Majoranas (Fig.~\ref{fig-setup} inset). In this setup, Majorana fermions define a nonlocal topological qubit, playing the role of a quantum spin-$1/2$ impurity interacting with the effective spin-$1$ of conduction electrons. This system therefore displays an overscreened single-channel Kondo effect with non-Fermi-liquid low energy behavior,\cite{Beri-Cooper-topological-2012} but with the overscreening that is stable even at low energies, unlike the usual overscreened multichannel case.\cite{nozieres1980kondo,affleck1990current,affleck1991kondo,affleck1991critical,affleck1992relevance,Oreg-Goldhaber-Gordon-two-2003,Potok-et-al-observation-2007,iftikhar2018tunable}

As in any Kondo system, important characteristics of strong correlations are revealed by the conduction electrons' spatial organization (i.e., the Kondo screening cloud\cite{SorensenPRB96,BarzykinPRL96,AffleckSimonPRL01,SimonAffleckPRB03,BordaPRB07,AffleckBordaSaleur08,Affleck-kondo-2010}) and their scattering properties. Here we focus on a quantity that provides information on both of these: the oscillating (as a function of position) part of the local electron tunneling density of states (tDOS), specifically its thermally smeared form motivated by scanning tunneling microscopy (STM). 
(For works focusing on the complementary nonoscillating part, see Ref.~\onlinecite{Eriksson-et-al-tunneling-2014,Agarwal-et-al-enhancement-2009}.) 
We will see that, unlike for free fermions, the amplitude of the oscillating tDOS component does not increase monotonically as temperature is lowered, but follows a form that depends on the position of the tunneling point relative to the Kondo cloud (see Fig.~\ref{fig-setup}). While the nonmonotonic temperature dependence is already indicative of strongly correlated scattering, strikingly, we also find that at the topological Kondo fixed point, in contrast even to correlated Fermi liquids,\cite{nozieres1974fermi,nozieres1980kondo} single-particle scattering becomes completely suppressed, and that this translates into the complete suppression of the oscillating part of the tDOS as the temperature (and bias voltage) tends to zero. Furthermore, for the minimal three-lead setup that we will focus on, the features that we predict can be turned off by decoupling any of the leads other than the one in which the tDOS is measured, which provides a \mbox{direct signature of the Majorana fermion nonlocality. }

\section{General Considerations}
\label{sec:generalcons}

We now turn to formulating the problem for our tDOS calculations. There are three terms that contribute to the Hamiltonian,\cite{Beri-Cooper-topological-2012,Galpin-et-al-conductance-2014}
\begin{equation} \label{Hamiltonian}
    \hat{H} = \hat{H}_{\text{leads}} + \hat{H}_c + \hat{H}_{\text{tun}}.
\end{equation}
The first term is due to the noninteracting, effectively spinless, conduction electrons in the three metallic leads (see Fig.~\ref{fig-setup}),
\begin{equation}
    \hat{H}_{\text{leads}} = \sum_{i = 1}^3 \int dk \, v_F k \, \hat{a}_{k, i}^\dagger \hat{a}_{k, i},
\end{equation}
where $v_F$ is the Fermi velocity. (The velocities can be taken to be the same for all leads without loss of generality.\cite{Chamon97}) We are working at sufficiently low energies, so that we can focus on the vicinity of Fermi wavenumber $k_F$ where the lead electron spectrum can be considered linear. The electron operator in momentum space $\hat{a}_{k, i}$ can be related to that in position space $\hat{\psi}_i(x)$ through
\begin{equation}
    \hat{\psi}_i(x) = \int dk \, \hat{a}_{k, i} \varphi_{k, i}(x),
\end{equation}
where $x \leq 0$ is the spatial coordinate in each lead such that the Majorana-lead junction is located at $x = 0$. The eigenfunction $\varphi_{k, i}(x)$ of the $i$-th lead (in the absence of Majorana-lead coupling) takes the form
\begin{equation}
    \varphi_{k, i}(x) = \frac{1}{\sqrt{2 \pi}} \left[ e^{i k_F x} \varphi_{k, i}^R(x) + e^{-i k_F x} \varphi_{k, i}^L(x) \right],
\end{equation}
where $\varphi_{k, i}^R(x) = e^{i k x}$ and $\varphi_{k, i}^L(x) = r_i e^{-i k x}$ are the right and left movers, respectively, and $r_i \equiv e^{i \theta_i}$ is the reflection amplitude of electrons at the lead endpoint.

Working at energies much below the induced superconducting gap, the superconducting island is characterized by the charging energy $E_c$ through the term
\begin{equation}
    \hat{H}_c = E_c (\hat{N} - q/e)^2,
\end{equation}
where $\hat{N}$ is the number operator of the electrons in the island, $q$ is the background charge, and $-e$ is the electron charge.\cite{Nazarov-Blanter-quantum-2009} The distance between any two Majorana zero modes is assumed to be large enough to ensure that the overlap of their localized wavefunctions can be ignored. In this case, the only tunneling mechanism we consider is when the electron of lead $i$ tunnels into the island through the Majorana $\hat{\gamma}_i$ with amplitude $t_i$, which is taken to be positive without loss of generality. The Hamiltonian for this is
\begin{equation}
    \hat{H}_{\text{tun}} = e^{i \hat{\phi}/2} \sum_{i = 1}^3 t_i \hat{\gamma}_i \hat{\psi}_i(0) + \text{h.c.},
\end{equation}
where $e^{\pm i \hat{\phi}/2}$ is an operator that changes the number of electrons in the island, $N \to N \pm 1$.\cite{Fu-electron-2010,Beri-Cooper-topological-2012}

We will be focusing on energy scales much smaller than $E_c$, so that the physics is dominated by virtual transitions connecting the lowest energy charge state to the neighboring ones. Focusing on the middle of the Coulomb blockade valley for simplicity (i.e., setting $q$ to be an integer multiple of $e$ in $\hat{H}_c$), this physics can be described by the effective Hamiltonian $\hat{H}_{\text{eff}} = \hat{H}_{\text{leads}} + \hat{H}_K$, where
\begin{equation}
    \hat{H}_K = \sum_{\alpha = 1}^3 g_\alpha \hat{I}_\alpha \otimes \hat{S}_\alpha,
\end{equation}
with $g_\alpha = \sum_{ij} |\epsilon_{\alpha ij}| {2 t_i t_j}/E_c$ where $\epsilon_{ijk}$ is the Levi-Civita matrix.\cite{Beri-Cooper-topological-2012,Beri-majorana-2013,Galpin-et-al-conductance-2014} This is the Kondo coupling mediating the interaction between the spin-$1/2$ topological qubit described by the operators \mbox{$\hat{S}_\alpha \equiv -\frac{i}{4} (\bm{\hat{\gamma}} \times \bm{\hat{\gamma}})_\alpha$} and the conduction electrons. The three lead species of the latter form an effective spin-$1$ density \mbox{$\hat{I}_\alpha = i \sum_{ij} \epsilon_{\alpha ji} \hat{\psi}_i^\dagger(0) \hat{\psi}_j(0)$}.\cite{Beri-Cooper-topological-2012}

As stated in the Introduction, we focus on the thermally smeared local tDOS of lead electrons, proportional to the STM differential conductance,\cite{Bruus-Flensberg-many-2004}
\begin{equation} \label{tDOS-A}
    \rho_i(x, V, T) \propto \int_{-\infty}^\infty d\omega \, \frac{\Do n_F(\omega - e V, T)}{\Do \omega} A_{ii}(x, \omega, T),
\end{equation}
where $n_F(\omega,T) = (1 + e^{\omega/T})^{-1}$ is the Fermi function, $V$ is the applied voltage between the STM tip and the lead, $T$ is the temperature, and $A_{ij}(x, \omega, T)$ is the electron spectral function. The latter can be calculated through the relation $A_{ij}(x, \omega, T) = -2 \, \text{Im}{[G_{ij}^R(x, x; \omega, T)]}$, where $G_{ij}^R(x, y; \omega, T)$ is the retarded Green's function, obtained from the Matsubara Green's function,
\begin{equation} \label{Matsubara}
    G_{ij}(x, y; i \omega_n, T) = -\int_0^{1/T} d\tau \, e^{i \omega_n \tau} \left\langle \T_\tau \left( \hat{\psi}_i(x, \tau) \hat{\psi}_j^\dagger(y, 0) \right) \right\rangle,
\end{equation}
with (imaginary) time-ordering operator $\T_\tau$, by performing an analytic continuation from the upper half plane to the real axis, $i \omega_n \to \omega + i \eta$.\cite{Bruus-Flensberg-many-2004} Note that at zero temperature, the tDOS is simply $\rho_i(x,V,0)\propto -A_{ii}(x, eV, 0)$. 

It is customary to introduce the so-called Kondo screening cloud, which is defined as the Kondo contribution to tDOS: If $\rho_{0, i}(x, V, T)$ denotes the tDOS of lead electrons when uncoupled from the Majoranas (which will be called the free cloud in the rest of the paper), then the Kondo cloud is defined by $\rho_{K, i}(x, V, T) \equiv \rho_i(x, V, T) - \rho_{0, i}(x, V, T)$.

The retarded Green's function $G_{ij}^R(x, x; \omega, T)$ can be written in terms of the $\T$-matrix $\T_{ij}(\omega,T)$, which is a key object encoding Kondo correlations,\cite{Affleck-Ludwig-exact-1993,Affleck-kondo-2010}
\begin{eqnarray} \label{Green-general}
    && G_{ij}^R(x, x; \omega, T) = G_{0, ij}^R(x, x; \omega) \\
    && \qquad + \, \sum_{kl} G_{0, ik}^R(x, 0; \omega) \T_{kl}(\omega, T) G_{0, lj}^R(0, x; \omega). \nonumber
\end{eqnarray}
Here $G_{0, ij}^R(x, y; \omega)$ is the retarded Green's function in the absence of interaction between electrons and Majoranas. This \textit{free} Green's function is obtained by performing the analytic continuation of Eq.~\eqref{Matsubara}, except that the averaging is performed by using the eigenstates of the noninteracting Hamiltonian $\hat{H}_{\text{leads}}$ instead.\cite{Affleck-Ludwig-exact-1993} It has the form
\begin{eqnarray}
    G_{0, ij}^R(x, x; \omega) &=& -\frac{i}{v_F} \delta_{ij} - \frac{i}{v_F} \delta_{ij} e^{i 2 K_i(\omega, x)}, \label{Green-free-x-x} \\
    G_{0, ij}^R(x, 0; \omega) &=& -\frac{2 i}{v_F} \cos{\left( {\theta_i}/2 \right)} \delta_{ij} e^{i K_i(\omega, x)}, \label{Green-free-x-0}
\end{eqnarray}
where
\begin{equation}\label{eq:Ki}
    2 K_i(\omega, x) \equiv 2 k_F |x| + \frac{2 \omega}{v_F} |x| + \theta_i.
\end{equation}
Due to the relation $\varphi_{k, i}^R(x) = r_i^\ast \varphi_{k, i}^L(-x)$ we have $G_{0, ij}^R(0, x; \omega) = G_{0, ij}^R(x, 0; \omega)$.\cite{Affleck-Ludwig-exact-1993}

A convenient parametrization of the $\T$-matrix is
\begin{equation} \label{T-matrix}
    \T_{ij}(\omega, T) = -\frac{i v_F}{4 \cos^2{\left( {\theta_i}/2 \right)}} \delta_{ij} \xi_i(\omega, T), \quad \theta_i \neq m\pi,
\end{equation}
where the $\delta_{ij}$ is present because only the diagonal terms survive the average over the impurity spin inherent in the definition of the Green's function. The angle $\theta_i = m\pi$ (with $m$ odd integer) is excluded because in this case the electron operator $\hat{\psi}_i(0)$ at the Majorana-lead junction vanishes, and therefore so does the Kondo coupling. 

Note that since $\T$ enters multiplicatively in Eq.~\eqref{Green-general}, the spatial features for $T = 0$ are entirely due to $G_0$ and $G_0^2$. To look for spatial features providing information on Kondo correlations we will therefore mostly focus on \mbox{$T > 0$} and for simplicity mostly take $V \to 0$. In this zero bias regime, the free cloud has a simple expression,
\begin{eqnarray} \label{free-cloud}
    && \rho_{0, i}(x, 0, T) \propto \\
    && \qquad \frac{2}{v_F} - \frac{4 \pi}{v_F} \frac{|x|}{L_T} \csch{\left( 2 \pi \frac{|x|}{L_T} \right)} \cos{\left( 2 k_F |x| + \theta_i \right)}, \nonumber
\end{eqnarray}
where $L_T \equiv {v_F}/T$ is the thermal length.

As exemplified by $G_{0, ij}^R$ and $\rho_{0, i}$ above, all quantities of our interest have two contributions: a nonoscillating and a $2k_F$-oscillating component. For the rest of the paper, we focus only on the latter because this is the piece that involves both ingoing and outgoing waves and thus encodes information about the scattering properties. [In contrast, at least for noninteracting lead electrons, the nonoscillating part is insensitive to the Kondo coupling as seen from Eqs.~\eqref{Green-general}-\eqref{Green-free-x-0}.] Denoting the oscillating components by the subscript $2k_F$, we have
\begin{equation} \label{Green}
    G_{2 k_F, ij}^R(x, x; \omega, T) = -\frac{i}{v_F} \delta_{ij} \left[ 1 - \xi_i(\omega, T) \right] e^{i 2 K_i(\omega, x)}.
\end{equation}

A key step in calculating the tDOS $\rho_i(x, V, T)$ is thus the calculation of $\xi_i(\omega, T)$. In the following, we will focus on two complementary regimes: high energies [$\max{(eV, T)} \gg T_K$] and low energies [$\max{(eV, T)} \ll T_K$], where $T_K$ is the Kondo temperature [see Sec.~\ref{sec-high-energy} below], considering mostly the zero bias case. These two regimes correspond to the vicinity of two renormalization group (RG) fixed points around which a perturbation theory can be developed: the free electron fixed point ($g_\alpha = 0$) for high energies and the topological Kondo fixed point for low energies. In the high energy regime, the function $\xi_i(\omega, T)$ will be obtained using perturbation theory in the Kondo couplings $g_\alpha$, considering terms up to third order. For low energies, we will adapt conformal field theory (CFT) results from Ref.~\onlinecite{Affleck-Ludwig-exact-1993} to our model.

\section{$2 k_F$-tDOS at High Energies}

\label{sec-high-energy}

After performing perturbation theory up to third order in the Kondo couplings, one finds
\begin{equation} \label{highxi1}
    \xi_i(\omega, T) = \pi^2 \left[ \widetilde{\lambda}_i^2 - 2 v_F \lambda_1 \lambda_2 \lambda_3 \int_{-\Lambda}^\Lambda dp \, \frac{\tanh{\left( \frac{v_F p}{2 T} \right)}}{\omega - v_F p + i \eta} \right],
\end{equation}
with cutoff $\Lambda$ and dimensionless couplings
\begin{equation}
    \lambda_i \equiv \frac{2 g_i}{\pi v_F} \frac{\Pi_j \cos{({\theta_j}/2)}}{\cos{({\theta_i}/2)}}, \qquad \widetilde{\lambda}_i^2 \equiv \sum_j \lambda_j^2 - \lambda_i^2.\label{eq:couplingdefs}
\end{equation}
For $v_F \Lambda \gg \omega, T$, Eq.~\eqref{highxi1} can be approximated as\cite{Cheung-Mattuck-removing-1970}
\begin{equation} \label{highxi2}
    \xi_i(\omega, T) \approx \pi^2 \left[ \widetilde{\lambda}_i^2 + 4 \lambda_1 \lambda_2 \lambda_3 \ln{\left( \frac{v_F \Lambda}{\sqrt{\omega^2 + 4 T^2}} \right)} \right].
\end{equation}
To gain some insight into the behavior of $\xi_i$, one may conveniently recast Eq.~\eqref{highxi2} using the weak coupling RG flow\cite{Beri-Cooper-topological-2012,Altland-Egger-multiterminal-2013,Beri-majorana-2013,Galpin-et-al-conductance-2014} $\frac{d\lambda_1}{d(\ln{E})} = -\lambda_2\lambda_3$ (and its cyclic permutations). Up to $O(\lambda_j^4)$, i.e., to the accuracy of our perturbation expansion, one finds
\begin{equation}
\xi_i(\omega, T)\approx \pi^2 \widetilde{\lambda}_i^2 (\sqrt{\omega^2 + 4 T^2}),
\end{equation}
where $\widetilde{\lambda}_i^2(E)$ is as in Eq.~\eqref{eq:couplingdefs}, but now in terms of the running couplings $\lambda_i(E)$. The latter satisfy \mbox{$\lambda_i(E)\sim [\ln(E/T_K)]^{-1}$}, where $T_K$ is the Kondo temperature, the single, emergent, energy scale characterizing the system and separating high and low energies. To the accuracy of our RG equations, it is given by\cite{Beri-Cooper-topological-2012,Altland-Egger-multiterminal-2013,Beri-majorana-2013,Galpin-et-al-conductance-2014} $T_K \sim v_F \Lambda e^{-1/{\bar\lambda}}$, where $\bar\lambda$ is the typical (bare) value of the $\lambda_j$ couplings. (We have $T_K = v_F \Lambda e^{-1/{\lambda_0}}$ in the isotropic $\lambda_i \equiv \lambda_0$ case.) 
As the weak coupling RG equations are the same as those for the conventional Kondo effect, it follows that the high-energy expression Eq.~\eqref{highxi2}, apart from an overall factor, is the same as that for ordinary single-channel Kondo systems,\cite{Affleck-kondo-2010} which can be represented by the lead-dot model to be discussed later (see Fig.~\ref{fig-models}a).

\begin{figure}[t]
    \centering
    \includegraphics[width=0.45\textwidth]{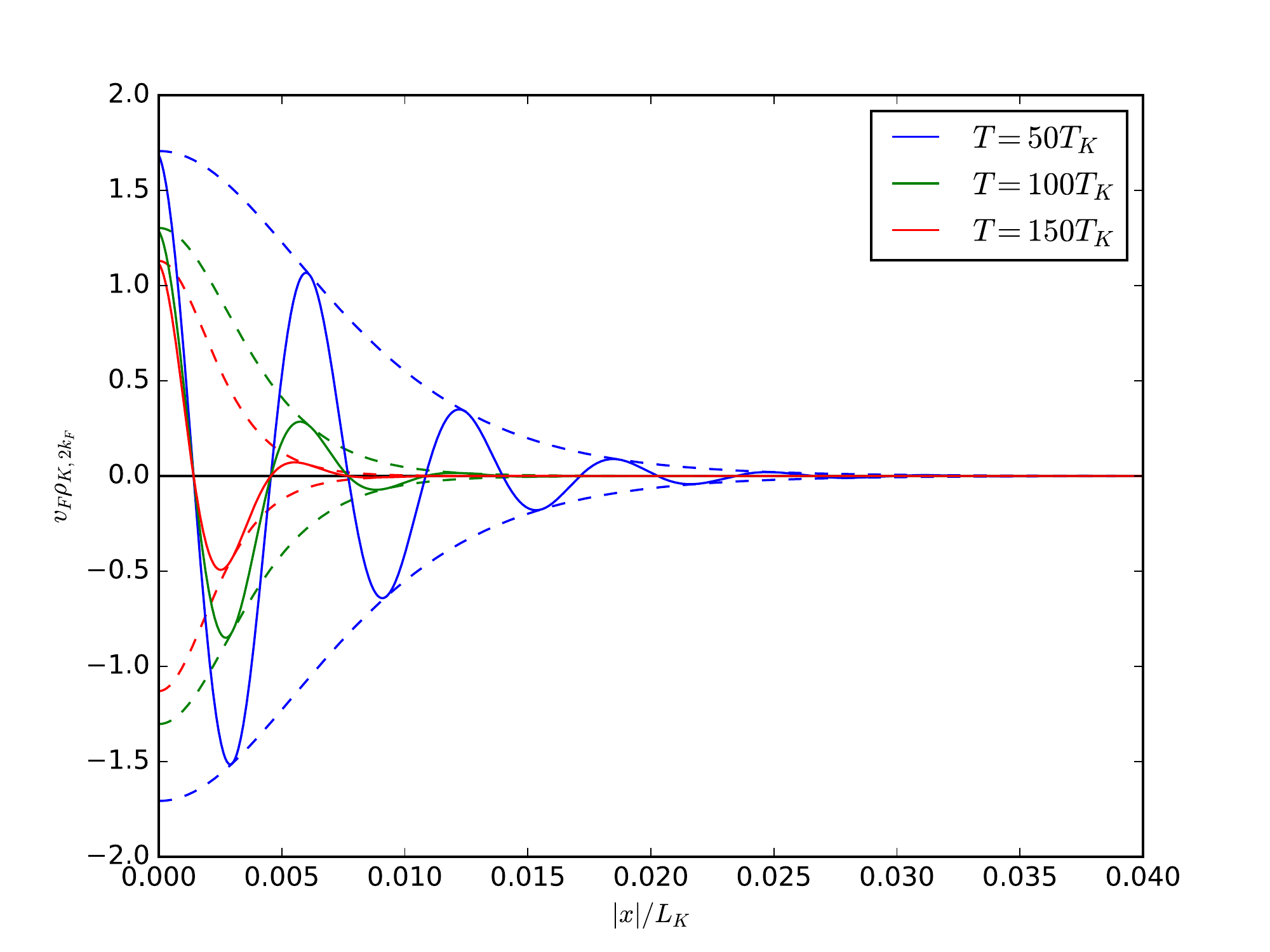}
    \caption{The Kondo cloud $\rho_{K, 2 k_F}$ (solid) and its amplitude $\widetilde{\rho}_{K, 2 k_F}$ (dashed) in the high energy regime, with $\theta = \pi/20$, $V = 0$, and $k_F L_K = 500$, where $L_K = {v_F}/{T_K}$ is the Kondo length. $\widetilde{\rho}_{K, 2 k_F}$ decays exponentially along $x$ on the scale of $L_T$. This high energy Kondo cloud profile, up to an overall factor, is the same as for ordinary single-channel Kondo systems represented by the lead-dot model (Fig.~\ref{fig-models}a).}
    \label{fig-KondoCloudHigh-var-T}
\end{figure}

Using our results for $\xi_i(\omega, T)$, we now discuss the high-energy form of the Kondo cloud and the oscillating tDOS. The essential features are already captured by the simplest, isotropic, case which we focus on henceforth. The Kondo cloud $\rho_{K, 2 k_F}$ at zero bias is given by
\begin{eqnarray}
    \rho_{K, 2 k_F} &\propto& \frac{8 \pi^3}{v_F} \frac{1}{\ln^2{\left( {2 T}/{T_K} \right)}} \frac{|x|}{L_T} \csch{\left( 2 \pi \frac{|x|}{L_T} \right)} \nonumber \\
    && \times \cos{\left( 2 k_F |x| + \theta \right)} \label{Kondo-cloud}
\end{eqnarray}
for $T \gg T_K$ (see the Appendix for details). This is depicted in Fig.~\ref{fig-KondoCloudHigh-var-T} for various temperatures as a function of ${|x|}/{L_K}$, where $L_K \equiv {v_F}/{T_K}$ is the Kondo length. 
Also shown is the amplitude of $\rho_{K, 2 k_F}$, denoted as $\widetilde{\rho}_{K, 2 k_F}$. (The temperatures used in Fig.~\ref{fig-KondoCloudHigh-var-T} may strictly be somewhat outside the domain where perturbation theory is accurate,\cite{Galpin-et-al-conductance-2014} but we believe that apart from overestimating the amplitude, the graph captures well the behavior.)
The exponential suppression of the function $({|x|}/{L_T}) \csch{\left( 2 \pi {|x|}/{L_T} \right)}$ in the limit $L_T \ll |x|$ is manifested in two complementary aspects: for fixed $T$, it shows that $\widetilde{\rho}_{K, 2 k_F}$ decays exponentially with increasing $|x|$ on the scale of $L_T$, and, for fixed $x$, given that $\rho_{0, 2 k_F}$ in Eq.~\eqref{free-cloud} contains the same function, one sees that it governs the exponential decay of the high temperature tail of the tDOS amplitude $\widetilde{\rho}_{2 k_F}$ shown in Fig.~\ref{fig-setup}. (The function $2\pi z\,\text{csch}(2\pi z)$ itself, describing both the envelope of the Kondo cloud apart from the logarithmic factor and the envelope for the $2k_F$ free cloud, is shown as a function of $z=|x|/L_T$ in Fig.~\ref{fig-tDOSLow-var-T} bottom panel,  with dash-dotted.)

The temperature dependence of $\widetilde{\rho}_{K, 2 k_F}$ and $\widetilde{\rho}_{2 k_F}$ is particularly interesting when there is a good scale separation so that $0 < |x| \ll L_K$. In this case,  the high energy regime $L_T \ll L_K$ displays a crossover upon lowering the temperature, from the exponential behavior discussed above for $L_T \ll |x|$ to a $|x| \ll L_T \ll L_K$ regime where $\widetilde{\rho}_{K, 2 k_F} \sim {1}/{\ln^2{({2 T}/{T_K})}}$ is governed by the Kondo logarithm. Extrapolating our results to $T \lesssim T_K$ beyond the perturbative regime we expect this logarithm-like increase of the Kondo cloud to develop into a contribution comparable to the free cloud and thus to govern the behaviour of $\widetilde{\rho}_{2 k_F}$ itself. Conversely, for $|x| \gg L_K$, the Kondo cloud and $\widetilde{\rho}_{2 k_F}$ remain exponentially suppressed even for $T \lesssim T_K$, thus the high temperature regime crosses over to the low temperature one without an intermediate logarithmic behavior. The temperature dependence of $\widetilde{\rho}_{2 k_F}$ in these two complementary regimes is shown in Fig.~\ref{fig-setup}. The presence versus the suppression of the logarithmic contribution may be used to estimate $L_K$, that is, the extent of the Kondo screening cloud.

\section{$2 k_F$-tDOS at Low Energies}

\label{sec-low-energy}

Now we turn to energies much below $T_K$. In this regime, weak $g_\alpha$ perturbation theory is inapplicable. Instead, we will work in the vicinity of the topological Kondo fixed point\cite{Beri-Cooper-topological-2012} and adapt the CFT results of Ref.~\onlinecite{Affleck-Ludwig-exact-1993} to obtain $\xi_i$. At the Kondo fixed point, i.e., at zero energy where the RG-irrelevant perturbations near this fixed point completely decayed, we have
\begin{equation} \label{T-matrix-without-perturbation}
   \xi_i(\omega\rightarrow0,T\rightarrow 0) = 1 - S_{(1)},
\end{equation}
where $S_{(1)}$ is the single-particle-to-single-particle scattering amplitude at the Fermi energy. It is given by
\begin{equation}
    S_{(1)} = \frac{{S_s^j}/{S_0^j}}{{S_s^0}/{S_0^0}},
\end{equation}
with $S_s^j = \sqrt{2/{(2 + k)}} \sin{[{\pi (2 j + 1) (2 s + 1)}/{(2 + k)}]}$, where $k$ is the level of the SU$(2)_k$ current algebra, $j$ is the spin of conduction electrons, and $s$ is the impurity spin.\cite{Affleck-Ludwig-exact-1993} For our model, $s = 1/2$, $j = 1$, and $k = 4$,\cite{Beri-Cooper-topological-2012,Galpin-et-al-conductance-2014} and therefore $S_{(1)} = 0$. This remarkable result signifies that, in stark contrast to Fermi liquid behavior, there is no single-particle scattering in topological Kondo systems at the Kondo fixed point. In terms of $\rho_{2k_F}$, which measures the single electron interference of incoming and outgoing waves, the vanishing of $S_{(1)}$ at the Kondo fixed point translates into $\rho_{2k_F} \to 0$ as $T,V \to 0$. The \mbox{$2 k_F$-tDOS} thus may be used to directly demonstrate the absence of single-particle scattering in the topological Kondo effect.

\begin{figure}[t]
    \centering
    \includegraphics[width=0.45\textwidth]{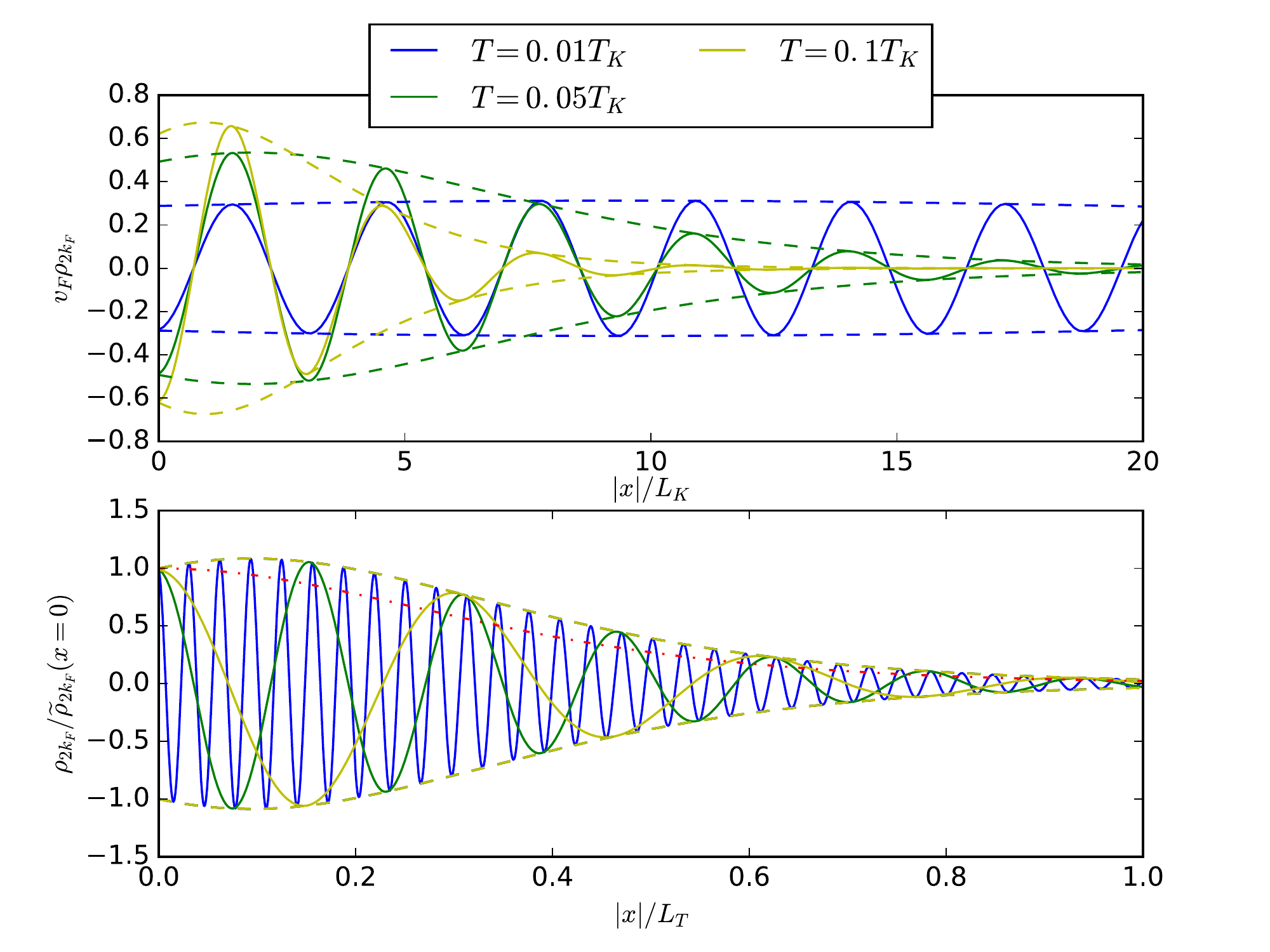}
    \caption{Top panel: The oscillating part of the tDOS $\rho_{2k_F}$
    (solid) and its amplitude $\widetilde{\rho}_{2 k_F}$ (dashed) in the low energy regime, with $\delta\lambda \! = \!-0.093$, $\theta \!=\! \pi/20$, $V \!=\! 0$ and  $k_F L_K \!=\! 1$.
    Bottom panel: Rescaling $\rho_{2k_F}$ and the corresponding scaling collapse of $\widetilde{\rho}_{2 k_F}$, as follows from \mbox{$\widetilde{\rho}_{2 k_F}\propto (T/T_K)^{1/3}h(|x|/L_T)$}. The universal scaling function $h(|x|/L_T)$ (dashed) is also compared to that of free fermions (dash-dotted),  \mbox{$h_\text{free}(|x|/L_T)=v_F\widetilde{\rho}_{0,2 k_F}/2$}. Note that the latter displays scaling collapse \emph{without} amplitude rescaling.
    }     
    \label{fig-tDOSLow-var-T}
\end{figure}

At low energies, but away from the Kondo fixed point, RG-irrelevant perturbations have to be taken into account and these lead to corrections $\delta \xi_i(\omega,T)$ to the function $\xi_i(\omega,T)$. For the neighborhood of an SU$(2)_k$ non-Fermi liquid Kondo fixed point, considering only the perturbation of the smallest scaling dimension $\Delta_s=1+2/(2+k)$ (the leading irrelevant operator), the CFT calculation of Ref.~\onlinecite{Affleck-Ludwig-exact-1993} shows that \mbox{$\delta \xi_i(\omega,T)\propto(T/T_K)^{\Delta_s-1}p(\omega/T,k)$} where $p$ is an integral expression. (Note that in contrast to the description in the high energy regime, the CFT does not determine $T_K$, but it rather enters as a microscopic parameter: it is the high energy cutoff of the low energy theory.) For the details of the calculation and the resulting general form of $p$ we refer the reader to Ref.~\onlinecite{Affleck-Ludwig-exact-1993}; here we only use the result specialized for the $k=4$ case of the three-lead topological Kondo effect.\cite{Beri-Cooper-topological-2012} We have
\begin{equation} \label{xi-low}
    \xi_i(\omega,T) = 1 - \delta\lambda \sqrt{3} (2 \pi {T}/{T_K})^{1/3} I(\omega,T).
\end{equation}
Here $\delta\lambda$ is proportional to the dimensionless coupling of the leading irrelevant operator and
\begin{eqnarray}
    I(\omega,T) &\equiv& \int_0^1 du \, \Bigg[ u^{-{i \omega}/{2 \pi T}} u^{-1/2} (1 - u)^{1/3} F(u) \nonumber \\
    && - \, \frac{\Gamma(5/3)}{\Gamma^2(4/3)} u^{-2/3} (1 - u)^{-4/3} \Bigg],
\end{eqnarray}
where $\Gamma$ is the gamma function and $F(u) \equiv {}_2F_1(4/3,4/3;1;u)$ is the hypergeometric function. We emphasize that the power law $T^{\Delta_s - 1} = T^{1/3}$ in Eq.~\eqref{xi-low} directly informs on the the scaling dimension $\Delta_s=4/3$.

For low energies, it is useful to consider two complementary regimes: when $T = 0$ but $V\neq 0$,  and when $V = 0$ with $T\geq0$. Though as mentioned earlier, the $T = 0$ spatial correlations are due to the free Green's functions, there is useful information to be obtained from $\xi$ and the overall amplitude of $\rho_{2k_F}$. For $\xi$ we find 
\begin{equation}\label{eq:tmatrix-T0}
    \xi_i(\omega,0) = 1 + \delta\lambda' \left[ \sqrt{3} - i \epsilon(\omega) \right] |{\omega}/{T_K}|^{1/3},
\end{equation}
where $\delta\lambda' \approx 3.98\, \delta\lambda$ and $\epsilon(\omega)$ is the sign function. This expression, firstly, may be used to specify $\delta\lambda$: while this is a free parameter for the CFT, the fact that the $\mathcal T$-matrix is a universal function\cite{Galpin-et-al-conductance-2014} of $\omega/T_K$ implies that $\delta\lambda$ also has a universal value. We can approximately obtain this by using Eq.~\eqref{eq:tmatrix-T0} to fit to the numerically exact results of Ref.~\onlinecite{Galpin-et-al-conductance-2014};  this gives $\delta\lambda\approx -0.093$. It also follows that for $T\rightarrow 0$, $\rho_{2 k_F}$ is a simple expression set by the second term in Eq.~\eqref{eq:tmatrix-T0}:  
\begin{eqnarray} \label{tDOS-zero-T}
    && \rho_{2 k_F}(x, V, 0) \propto \frac{2}{v_F} \delta\lambda' |{e V}/{T_K}|^{1/3} \\
    && \qquad \times \left\{ \sqrt{3} \cos{[2 K(e V, x)]} + \epsilon(e V) \sin{[2 K(e V, x)]} \right\}. \nonumber
\end{eqnarray}

In the $V = 0$, $T\geq0$ case, we plot the oscillating component of the tDOS of lead electrons for various temperatures (top panel of Fig.~\ref{fig-tDOSLow-var-T}).
As the function of temperature, $\rho_{2k_F}$ is gradually suppressed for all $x$ as $T$ decreases. This is in contrast to the free cloud which gradually saturates upon lowering the temperature [see Eq.~\eqref{free-cloud} and Fig.~\ref{fig-setup}]. 
Note that at low energies, since $L_K$ became the short distance cutoff (as follows from $T_K$ being the high energy cutoff), the only length scale that can set long distance features is the thermal length $L_T$. This can be made manifest by noting that the tDOS amplitude $\widetilde{\rho}_{2 k_F}$, as shown in the Appendix, \text{admits the scaling form} $\widetilde{\rho}_{2 k_F}\propto (T/T_K)^{1/3}h(|x|/L_T)$ with the universal scaling function 
$h(|x|/L_T)=\widetilde{\rho}_{2 k_F}/\widetilde{\rho}_{2 k_F}(x\!\rightarrow\!0)$. 
The corresponding scaling collapse, illustrated in the bottom panel of Fig.~\ref{fig-tDOSLow-var-T}, may serve as a useful characteristic of the spatial organization of conduction electrons near the topological Kondo fixed point, and as means to demonstrate the $T^{1/3}$ law (and thus the scaling dimension $\Delta_s$) governing the suppression of the oscillations as the temperature is lowered (Fig.~\ref{fig-setup}). [Extracting $\widetilde{\rho}_{2 k_F}$ and thus the scaling function $h(|x|/L_T)$ from ${\rho}_{2 k_F}$ in practice may be facilitated by oscillation extrema much denser than $L_T$, including $k_FL_K>1$. The latter is not inconsistent with $L_K$ being the short distance cutoff of the CFT, since that only limits the spatial resolution for $\widetilde{\rho}_{2 k_F}$, and not the wavelength of the $2k_F$ oscillations.] We note that a similar form, $\widetilde{\rho}_{2 k_F}=f(T/T_K)h_\text{free}(|x|/L_T)$, holds also in the high energy regime. The scaling function  in that case, $h_\text{free}(z=|x|/L_T)=2\pi z\text{csch}(2\pi z)$, is however the same as for free fermions and thus unlike $h(|x|/L_T)$ for low energies, does not provide information on Kondo features. The two curves are contrasted in the bottom panel of Fig.~\ref{fig-tDOSLow-var-T}.

\section{Discussion and Conclusions}

A common feature shared by our high- and low-temperature results is the thermal-length-controlled large-$|x|$ decay of the amplitude $\widetilde{\rho}_{2 k_F}$ [see Eqs.~\eqref{free-cloud} and \eqref{Kondo-cloud} and Fig.~\ref{fig-tDOSLow-var-T}]. In terms of the temperature dependence of $\widetilde{\rho}_{2 k_F}(T,|x|)$ (illustrated in Fig.~\ref{fig-setup}), the role of $|x|$ is thus to control the competition between the thermal and Kondo lengths $L_T$ and $L_K$ by setting the low temperature cutoff above which $L_T$ dominates.

Considering that our high- and low-energy asymptotics are expected to be accurate\cite{Galpin-et-al-conductance-2014}
for $T\gtrsim10^{3}T_K$ and $T\lesssim 10^{-2}T_K$, respectively, one needs slightly exaggerated $|x|/L_K$ values to achieve good scale separation while staying within the strict domain of validity of our theory. (Fig.~\ref{fig-setup} uses \mbox{$|x|=10^{-5}L_K$} for $|x|\ll L_K$ and $|x|=200L_K$ for \mbox{$|x|\gg L_K$}.) We however believe that the behavior is captured qualitatively correctly also for less conservative values of $T$, as in Figs.~\ref{fig-KondoCloudHigh-var-T}, \ref{fig-tDOSLow-var-T}, which allows for scale separation for more moderate $|x|/L_K$. It would be interesting to compare this expectation to results from numerical renormalization group calculations of the $\mathcal{T}$-matrix which are valid also between the asymptotic regimes.\cite{Mitchell2011} 

To work in the regime dominated by topological Kondo physics, as we noted in Sec.~\ref{sec:generalcons}, temperature and voltage should be much smaller than the induced superconducting gap $\Delta$ and the charging energy $E_c$. For topological Kondo setups, these are expected to be comparable to those in recent Al-InAs nanowire devices\cite{Albrecht-et-al-exponential-2016,Albrecht-et-al-qp-2017,Sesoft2018} where \text{$E_c\sim\Delta\sim 0.1\text{meV}$}. These values also provide an estimate for the energy window within which to set the Kondo temperature $T_K$ using suitable tunnel couplings. 
For clear $2k_F$ oscillations, low disorder leads with mean free path $l$ satisfying $k_F l\gg 1$ are advantageous, as in recent ballistic InSb nanowire\cite{zhang2017ballistic} and InAs 2DEG based\cite{JSLee2017} devices with $l\sim 1\mu\text{m}$, considering a typical Fermi wavelength\cite{Jespersen09} of $\lambda_F\sim 20\text{nm}$. In terms of the oscillation amplitude itself, as shown in Fig.~\ref{fig-setup}, these are appreciable: the maximal amplitude (as the function of $T$), even for $|x|\!=\!200L_K$, is just an order of magnitude smaller than that of the saturated ($T=0$) free $2k_F$ tDOS,  and it increases with decreasing $|x|$. Provided that the free $2k_F$ tDOS is accessible in the device components forming the leads, the features we predict should also be visible.

\begin{figure}[t]
    \centering
    \includegraphics[width=0.4\textwidth]{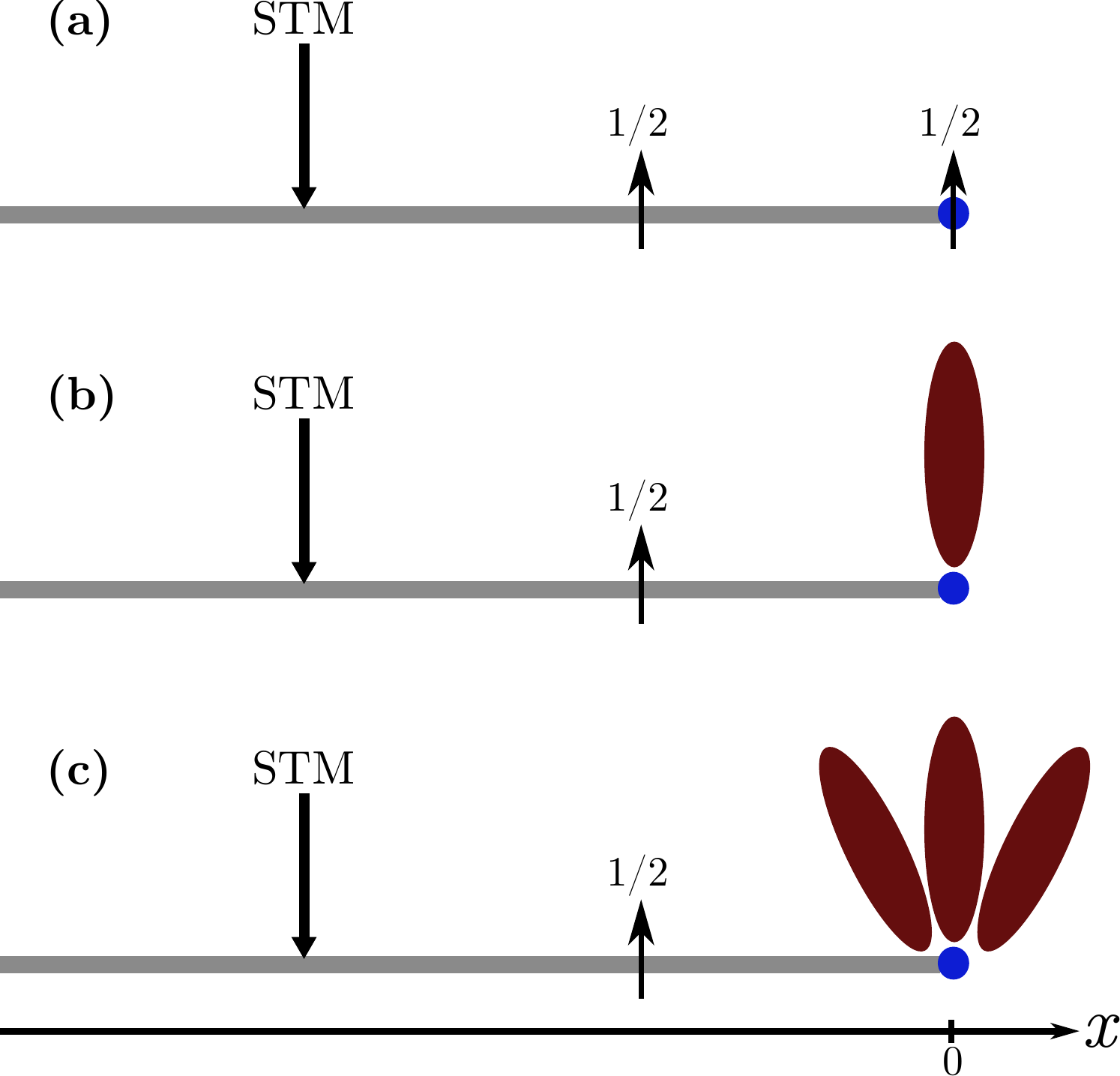}
    \caption{(a) The lead-dot model, with a small quantum dot forming a spin-$1/2$ ``impurity" coupled to a lead of spin-$1/2$ conduction electrons. (b) The modified Oreg-Goldhaber-Gordon model,\cite{Oreg-Goldhaber-Gordon-two-2003} where now the small quantum dot is also coupled to a large quantum dot which acts as a reservoir of spin-$1/2$ electrons. This model is a modified version of the one confirmed experimentally to host the two-channel Kondo (2CK) effect.\cite{Potok-et-al-observation-2007} (c) The generalized Oreg-Goldhaber-Gordon model, with a small dot and three large dots. In order for this model to host the four-channel Kondo (4CK) effect, the couplings to the large dots and the lead have to be symmetric.}
    \label{fig-models}
\end{figure}

While in Sec.~\ref{sec-high-energy} we found that the high energy behavior of $\widetilde{\rho}_{2 k_F}$ and ${\rho}_{2 k_F}$ is similar to that in more conventional Kondo systems, there are important differences in the low energy regime. It is thus useful to contrast our results to such more conventional, single- and multi-channel Kondo systems. The simplest, single-channel Kondo effect arises in the lead-dot model shown Fig.~\ref{fig-models}a. Here $S_{(1)} = -1$, corresponding to a $\pi/2$ phase shift in single-particle scattering.\cite{nozieres1974fermi,Affleck-Ludwig-exact-1993} This system is a local Fermi liquid at low energies. The low temperature $\widetilde{\rho}_{2 k_F}$ thus is similar to the free cloud, increasing upon lowering temperature as the reduction of thermal smearing allows more and more single-particle interference.

Our findings are also in contrast with two-channel Kondo (2CK) systems proposed and later experimentally studied by Oreg and Goldhaber-Gordon.\cite{Oreg-Goldhaber-Gordon-two-2003,Potok-et-al-observation-2007} Their system is a two-lead 2CK model where there is a linear combination of modes without single-particle scattering at the Fermi energy [i.e., $S_{(1)}=0$ for this linear combination], but there is another linear combination which still has single-particle scattering, translating into ${\rho}_{2 k_F}\neq 0$ at the 2CK fixed point.\cite{Carmi-et-al-transmission-2012} It is interesting to note, however, that one may modify this model by removing one of the leads while maintaining the coupling symmetry between the remaining lead and the large dot (leading to the setup in Fig.~\ref{fig-models}b). Now there is only the $S_{(1)}=0$ mode, which leads to \mbox{${\rho}_{2 k_F}= 0$} at the 2CK fixed point. However, as temperature is lowered, the power-law suppression is different: $\widetilde{\rho}_{2 k_F}(x, 0, T) \sim T^{1/2}$, as can be shown by adapting our considerations to this case.

A system for which we do find the same power law (and SU$(2)_4$ current algebra\cite{Fabrizio-Gogolin-toulouse-1994}) as for the topological Kondo effect is the 4CK model, corresponding to the generalized Oreg-Goldhaber-Gordon setup with three large dots\cite{Oreg-Goldhaber-Gordon-two-2003,Florens-Rosch-climbing-2004} (Fig.~\ref{fig-models}c). However, now the conduction electrons have $j = 1/2$, which results in $S_{(1)} = 1/{\sqrt{3}}$ and thus ${\rho}_{2 k_F}\neq 0$ at the 4CK fixed point.

Generalizing our considerations, it is also interesting to note that one may use $\rho_{2 k_F}$ at zero temperature and bias to measure $|S_{(1)}|$ in a range of Kondo and other quantum impurity systems, provided there is only one value of $k_F$ to consider (as is the case for single channel leads). To this end, one takes the ratio between $\widetilde{\rho}_{2 k_F} \propto{2 |S_{(1)}|}/{v_F}$ and the nonoscillating tDOS component ${\rho}_{k=0} \propto 2/{v_F}$. This is useful since the proportionality factor [originating from Eq.~\eqref{tDOS-A}] is the same in both cases and depends only on the density of states and physical characteristics of the STM tip.\cite{Bruus-Flensberg-many-2004,Gottlieb-Wesoloski-bardeens-2006} Therefore, $\widetilde{\rho}_{2 k_F}/{\rho}_{k=0}=|S_{(1)}|$. 

To summarize, we have shown that the oscillating component $\!\rho_{2 k_F}\!$ of the tDOS provides valuable novel insights into the strong correlations in the topological Kondo effect. At zero bias, the difference in the behavior of the amplitude $\!\widetilde{\rho}_{2 k_F}\!$ as a function of temperature for different values of $\!{|x|}/{L_K}\!$ provides information on the size of the Kondo length $L_K$, and hence the extent of the Kondo screening cloud. At low temperatures, $\widetilde{\rho}_{2 k_F}$ admits a universal scaling form $\widetilde{\rho}_{2 k_F}\!\propto\! (T/T_K)^{1/3}h(|x|/L_T)$ characterizing the conduction electrons' spatial organization near the topological Kondo fixed point. As temperature and bias tend to zero, $\rho_{2 k_F}$ becomes completely suppressed, revealing that in the topological Kondo effect, single-particle scattering is \mbox{entirely absent at the Fermi energy.}

\vspace*{1em}
\section*{Acknowledgements}

We thank S. Das and V. Dwivedi for input at early stages of the work, and R. Egger and M. Galpin for useful discussions. This research was supported by the Royal Society, the Indonesian Endowment Fund for Education (LPDP) scholarship, the EPSRC grant EP/M02444X/1, and the ERC Starting Grant No.~678795 TopInSy.

\section*{Appendix}

In this Appendix, we study the temperature dependence
of the amplitude of the $2 k_F$-tDOS with isotropic couplings at zero bias, both in the high and low energy regimes. If one denotes $y \equiv \omega/T$, in the high energy regime the spectral function contribution relevant for the Kondo cloud amplitude  has the form
\begin{eqnarray}
    && \widetilde{A}_{K, 2 k_F}{({|x|}/{L_T}, y, {T}/{T_K})} = \\
    && \qquad -\frac{4 \pi^2}{v_F} \frac{1}{\ln^2{\left( \frac{T}{T_K} \sqrt{y^2 + 4} \right)}} \cos{\left( 2 y \frac{|x|}{L_T}  \right)}. \nonumber
\end{eqnarray}
The Kondo cloud then can be obtained from Eq.~\eqref{tDOS-A}. For $T \gg T_K$, its amplitude can be approximated by the integral
\begin{eqnarray}
    && \widetilde{\rho}_{K, 2 k_F}{({|x|}/{L_T}, 0, {T}/{T_K})} \nonumber \\
    && \qquad \propto \frac{\pi^2}{v_F} \int_{-\infty}^\infty dy \, \frac{\sech^2{(y/2)}}{\ln^2{({2 T}/{T_K})}} \cos{\left( 2 y \frac{|x|}{L_T}  \right)} \nonumber \\
    && \qquad = \frac{8 \pi^3}{v_F} \frac{1}{\ln^2{({2 T}/{T_K})}} \frac{|x|}{L_T} \csch{\left( 2 \pi \frac{|x|}{L_T} \right)}. \nonumber \\
\end{eqnarray}
Since $\frac{|x|}{L_T} \csch{\left( 2 \pi \frac{|x|}{L_T} \right)} \to 2 \frac{|x|}{L_T} e^{-2 \pi {|x|}/{L_T}}$ for $L_T \ll |x|$, this shows that the amplitude of the Kondo cloud decays exponentially with $|x|$ on the scale $L_T$.

In the low energy regime, one has
\begin{eqnarray}
    && \widetilde{A}_{2 k_F}{({|x|}/{L_T}, y, T)} = \\
    && \qquad \frac{2}{v_F} \delta\lambda \sqrt{3} \left( 2 \pi \frac{T}{T_K} \right)^{1/3} \text{Re}{\left[ I(y) e^{i  2 y \frac{|x|}{L_T} } \right]}. \nonumber
\end{eqnarray}
Therefore, the amplitude of the $2 k_F$-tDOS is
\begin{equation}
    \widetilde{\rho}_{2 k_F}{({|x|}/{L_T}, 0, T)} \propto -\frac{2}{v_F} \delta\lambda \sqrt{3} \left( 2 \pi \frac{T}{T_K} \right)^{1/3} h{\left( \frac{|x|}{L_T} \right)},
\end{equation}
for some function $h$. To extract the $T\rightarrow0$ asymptotic power law, we may take $L_T\gg|x|$, and thus substitute $h{({|x|}/{L_T})} \sim h(0)$. This gives the $T^{1/3}$ decay.


%

\end{document}